\documentclass[twocolumn]{article}
\usepackage{graphicx}
\usepackage{bm}
\usepackage{hyperref}
\begin{document}

\title{Fast Non-destructive temperature measurement of two-electrons atoms in a magneto-optical trap}
%\subtitle{}
\author{Matteo Cristiani\thanks{The first two authors contributed equally to the present work}
\and Tristan Valenzuela${}^{\star}$
\and Hannes Gothe
\and J\"urgen Eschner
\\
\texttt{(mail: matteo.cristiani@icfo.es)}
\\
\\
ICFO-Institut de Ciencies Fotoniques,
\\
Mediterranean Technology Park,
\\
08860 Castelldefels (Barcelona), Spain
}
%\mail{matteo.cristiani@icfo.es}
%\institute{
%ICFO-Institut de Ciencies Fotoniques,
%Mediterranean Technology Park,
%08860 Castelldefels (Barcelona), Spain
%}

\date{Received: \today / Revised version: \today}

\maketitle

\bibliographystyle{unsrt}
\begin{abstract}
We extend the technique originally proposed by Honda et al. \cite{Honda1999} to measure the temperature of Ytterbium and alkine-earth atoms confined in a Magneto-Optical Trap (MOT). The method is based on the analysis of excitation spectra obtained by  probing the ${}^1\mathrm{S}_0\rightarrow{}^3\mathrm{P}_1$ inter-combination line. Thanks to a careful analysis and modeling of the effects caused by the MOT light on the probe transition we overcome the resolution and precision limits encountered in previous works \cite{Honda1999,Loftus2000a}. Ground state light shift and Rabi broadening are measured and successfully compared with calculated values. This knowledge allows us to properly extract the Doppler contribution to the linewidth, thus obtaining a reliable measurement of the cloud temperature. We finally show how spectroscopy on free-falling atoms provides an alternative method to determine the sample temperature which resembles the standard time-of-flight technique.\\
{\bf PACS:} 37.10.De, 37.10.Gh, 37.10.Vz, 32.10.-f
\end{abstract}

\bibliographystyle{apsrev}

\section{\label{sec:intro}Introduction}

Since the first experimental realization of magneto-optical cooling and trapping \cite{Phillips1998}, cold and ultracold atomic gases have become an ubiquitous tool in physics. Solid state physics and many-body physics \cite{Bloch2008}, metrology and quantum information \cite{Hughes2004} are just a few examples of research fields boosted by this new resource. Due to their favorable spectroscopic properties, alkali atoms have been considered as the most natural choice for experiments with ultracold vapours and they have played a prominent role in the spectacular advances which lead to the first observation of Bose-Einstein condensation in atomic gases \cite{Anderson1995,Bradley1995,Davis1995}. More recently, Ytterbium and two-electrons atoms in general have raised interest due to their possible application in metrology \cite{Takamoto2005} and quantum information \cite{Shibata2009}.

\begin{table}[h!]
\caption{\label{tab:table1}Characteristics of the ${}^1S_0\rightarrow{}^1P_1$ (trap) and ${}^1S_0\rightarrow{}^3P_1$ (probe) transitions. The Doppler temperature is defined as $T_d=\hbar \Gamma / 2 k_B$. The MOT capture range is $v_c=\Gamma/k$, with $k=2\pi/\lambda$. According to the convention used in the text all the quantities relative to the trap (probe) line are indicated by the subscript $t$ ($p$).}
\begin{center}
\begin{tabular}{l|ccccc}
Transition			& $\lambda$		& $\Gamma$			& $I_s$
& $T_d$			& $v_c$			\\
				& $\mathrm{nm}$	& $2\pi\,\times\,\mathrm{MHz}$	& $\mathrm{mW/cm^2}$
& $\mathrm{\mu K}$	& $\mathrm{m/s}$	\\
\hline
${}^1S_0\rightarrow{}^1P_1$	& $399$		& $29$				& $59$
& $695$			& $10$			\\
${}^1S_0\rightarrow{}^3P_1$ & $556$		& $0.182$			& $0.14$
& $4.5$			& $0.1$\\
\end{tabular}
\end{center}
\end{table}

\begin{figure}[h]
\includegraphics[width=0.3\textwidth]{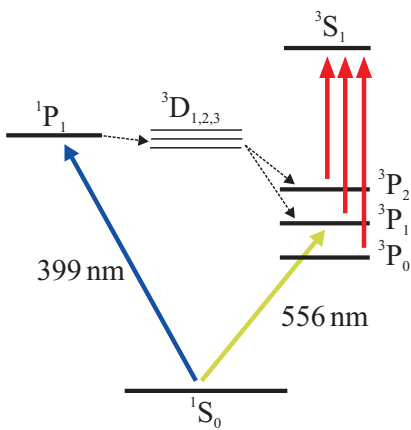}
\caption{\label{fig:levels}(Colors online). Scheme of the lowest energy states of Ytterbium. The ${}^1S_0\rightarrow{}^1P_1$ (${}^1S_0\rightarrow{}^3P_1$) transition is indicated by a blue (green) arrow.}
\end{figure}

Ytterbium atoms are characterized by a rich electronic structure. The scheme of figure \ref{fig:levels} shows that the ground state ${}^1S_0$ can be optically coupled to the ${}^1P_1$ and ${}^3P_1$ states. Largely different linewidths are associated with the two transitions (see table \ref{tab:table1}). This aspect is reflected in the different cooling regimes which can be achieved in the experiment. In the case of the ${}^1S_0\rightarrow{}^1P_1$ transition a relatively large capture velocity is associated with high Doppler temperature. On the other hand for the ${}^1S_0\rightarrow{}^3P_1$ transition a much lower temperature can be achieved at the expense of a narrower capture range. Cooling and trapping of $\mathrm{Yb}$ atoms has been successfully demonstrated using both transitions \cite{Honda1999,Kuwamoto1999,Loftus2001,Maruyama2003a,Park2003,Zhao2008}. Combining these techniques with evaporative cooling in optical dipole traps, quantum degeneracy has been reached \cite{Takasu2003a,Fukuhara2007,Fukuhara2009}.

Nevertheless it has been found that for alkaline-earth and Ytterbium atoms trapped by the ${}^1S_0\rightarrow{}^1P_1$ transition the measured cloud temperature is significantly higher than the value predicted by standard Doppler theory. Several explanations have been proposed to justify this discrepancy. Inelastic collisions \cite{piilo2004} and spatial intensity fluctuations \cite{Chaneliere2005} have been suggested as possible sources of extra heating. More recently a 3D anlysis of the MOT dynamics has been carried out \cite{Choi2008}. The predicted temperature is in good agreement with the experimental data for ${}^{88}\mathrm{Sr}$ \cite{Xu2002,Xu2003} when $I/I_{s} \le 1$. This approach seems to be a promising way to take into account the ``non-Doppler'' behavior seen in the experiments. However it is important to compare the model with data obtained from experiments performed with other alkaline-earth atoms.

In this context we extend the method originally proposed by Honda et al. \cite{Honda1999} to measure the cloud temperature in a non-destructive way by probing the weak ${}^1S_0\rightarrow{}^3P_1$ intercombination transition. As stressed by Loftus et al. \cite{Loftus2000a} this technique has great potential since it allows online monitoring of trap properties like the atomic cloud temperature and size, or the average magnetic field at the MOT position. In this previous work the large linewidth of the probe laser combined with the Rabi broadening induced by the trapping beam limited the authors to setting an upper bound for the temperature.

In our case, the relatively narrow linewidth of our $556\,\mathrm{nm}$ probe laser ($\approx2\pi\times10\,\mathrm{kHz}$) allows us to resolve the line shape of the excitation spectrum. From the ${}^1S_0\rightarrow{}^3P_1$ spectra we extract the Gaussian contribution to the line profile, thus measuring the sample temperature. Rabi broadening and light shift are also measured and their values are compared with the results of a simple model describing the ${}^1S_0\rightarrow{}^1P_1$ trapping transition. We then show how time-of-flight temperature measurement can be carried out with the same technique.

We find that the measured temperature is systematically above the standard Doppler limit, as observed in \cite{Xu2002,Xu2003}. We compare our results with the model developed in \cite{Choi2008}. We also discuss how this technique could give a better insight into the dynamics of trapping and cooling of two-electrons atoms.

In the last section, we present a measurement of the cloud temperature and size taken in only $1\,\mathrm{ms}$ without disturbing the MOT dynamics. This result puts in evidence that the method discussed in the present paper can be regarded as a tool to probe the ``instantaneous'' cloud properties in a non-distructive way. 

\begin{figure}[ht]
\includegraphics[width=0.45\textwidth]{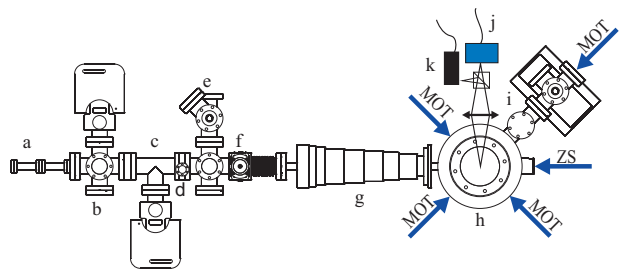}
\caption{\label{fig:apparatus}Schematic view of the vacuum chamber. (a) atomic beam oven; (b) atomic beam probe; (c) differential pumping stage; (d) shutter; (e) transverse cooling area (not used in the present work); (f) gate valve; (g) Zeeman slower; (h) main chamber; (i) ion and non-evaporative getter (NEG) pumps; (MOT) MOT beams; (ZS) Zeeman slower beam.}
\end{figure}

\section{\label{sec:method}Interpretation and analysis of excitation spectra}
Our apparatus is schematically represented in figure \ref{fig:apparatus}. In the main chamber a ${}^{174}\mathrm{Yb}$ magneto-optical trap (MOT) is operated on the (blue) ${}^1S_0\rightarrow{}^1P_1$ transition. The $399\,\mathrm{nm}$ source is a frequency-doubled diode laser stabilized to a hollow cathode lamp \cite{Corwin1998,Kim2003}. The MOT is constituted by three retroreflected beams ($10\,\mathrm{mm}$ diameter, $2.7\,\mathrm{mW}$ maximum power). About $10\,\mathrm{mW}$ are used for the Zeeman slower which decelerates thermal atoms from the oven ($500\mathrm{^{\circ}C}$) into the capture range of the MOT.

\begin{figure}[h]
\includegraphics[width=0.45\textwidth]{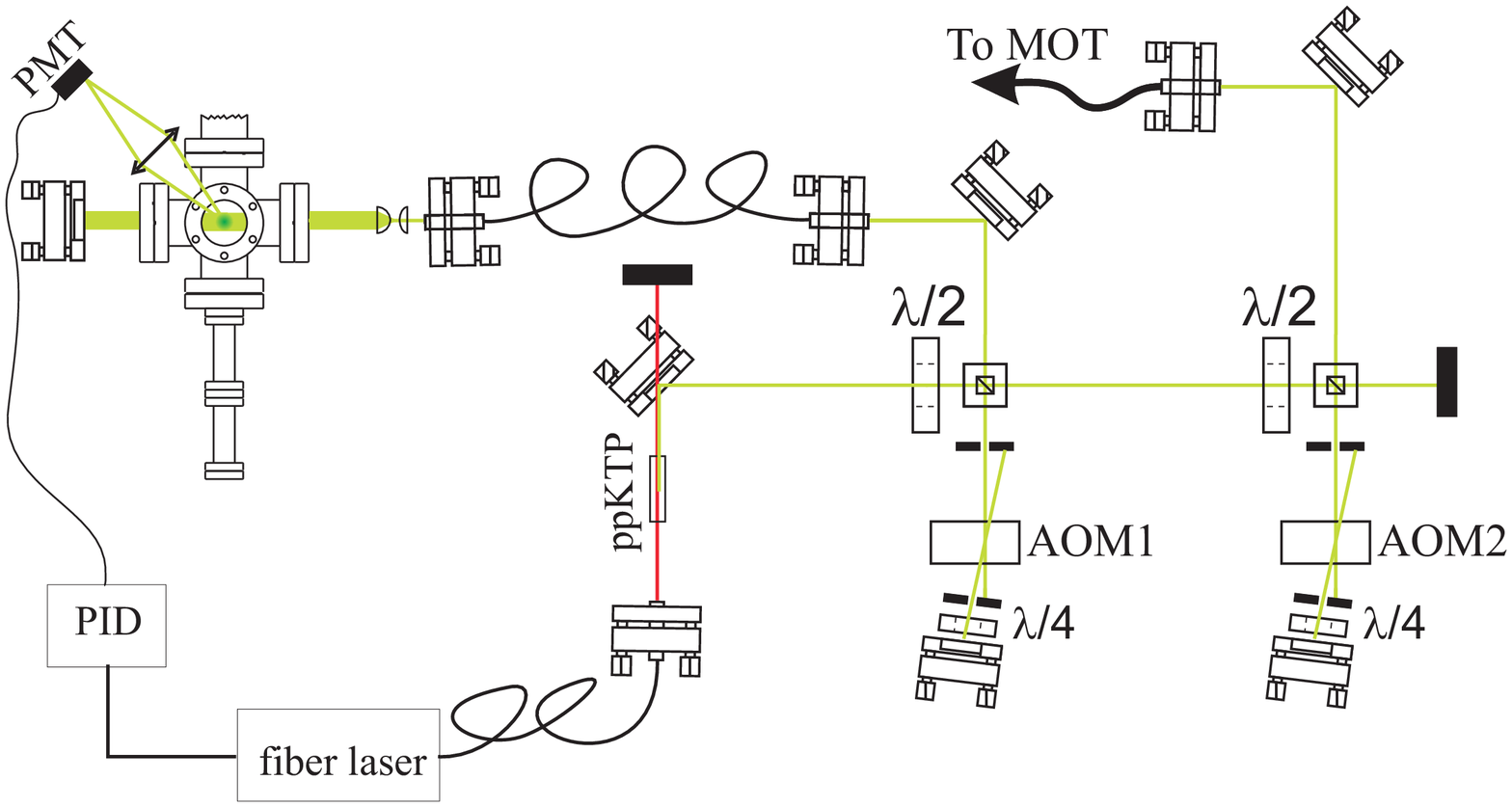}
\caption{\label{fig:greenlaser}Schematic view of the green laser setup. The $1112\,\mathrm{nm}$ IR light is frequency doubled by  single passage through a ppKTP crystal. Excitation spectroscopy on the atomic beam --- top left, see also figure \ref{fig:apparatus}, part b --- provides the error signal for locking the $556\,\mathrm{nm}$ frequency to the ${}^1S_0\rightarrow{}^3P_1$ transition. A double passage Acousto-Optic Modulator (AOM2) is used to vary the probe beam frequency.}
\end{figure}

While the atoms are confined in the blue MOT we excite the ${}^1S_0\rightarrow{}^3P_1$ transition using probe light overlapped with the MOT beams. Each probe beam has $40\,\mathrm{\mu W}$ power and $7\,\mathrm{mm}$ diameter. The $556\,\mathrm{nm}$ light is generated by frequency-doubling of a $1112\,\mathrm{nm}$ fiber laser in a ppKTP crystal. The probe linewidth of $2\pi\times10\,\mathrm{kHz}\sim\Gamma_p/10$ has been measured by analyzing the transmission of an High-Finesse ($\mathcal{F}=5\,\times\,10^4$) Fabry-Perot resonator (length $4.74\,\mathrm{cm}$ and linewidth $2\pi\times70\,\mathrm{kHz}$). The position of the resonance transition used as reference for the laser lock circuit, is determined within $2\pi\times200\,\mathrm{kHz}$. This sets the limit of our absolut frequency calibration. The green laser setup is detailed in figure \ref{fig:greenlaser}.

\begin{figure}[h]
\includegraphics[width=0.4\textwidth]{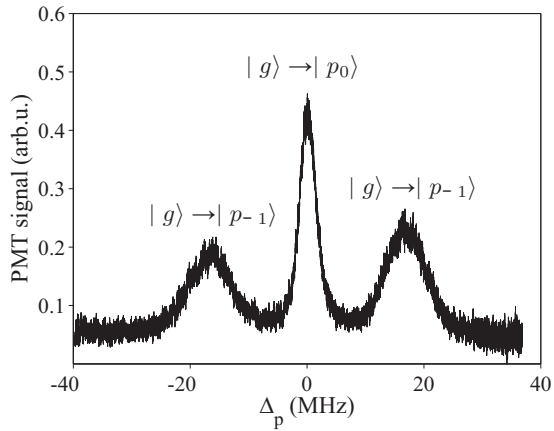}
\caption{\label{fig:typical}Typical excitation spectrum obtained by probing the ${}^1S_0\rightarrow{}^3P_1$ transition on $\mathrm{Yb}$ atoms trapped in a ${}^1S_0\rightarrow{}^1P_1$ MOT and measuring the fluorescence rate. The central peak is originated by the $\vert g \rangle\rightarrow\vert p_0 \rangle$ excitation process. The $\vert g \rangle\rightarrow\vert p_{\pm1} \rangle$ channels are responsible for the two side peaks. See figure \ref{fig:levels-LS} for a detailed description of the notation used in labeling the relevant energy levels.}
\end{figure}

\begin{figure}[ht]
\includegraphics[width=0.45\textwidth]{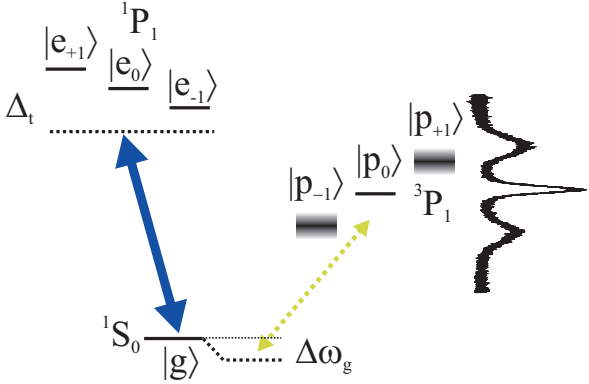}
\caption{\label{fig:levels-LS}(Colors online). Schematic representation of the energy levels involved in the ${}^1S_0\rightarrow{}^3P_1$ fluorescence measurement. $\vert g \rangle$ stands for the ${}^1S_0$ ground state, $\vert e_{0,\pm1} \rangle$ for the ${}^1P_1$ multiplet and $\vert p_{0,\pm1} \rangle$ stand for the $m=0,\pm1$ Zeeman sublevels of the ${}^3P_1$ state. The $\vert g \rangle\rightarrow\vert e \rangle$ trapping light (blue continuous arrow) induces a light shift on the $\vert g \rangle$ ground state. The probe beam (green dashed arrow) excites the $\vert g \rangle \rightarrow \vert p_{0,\pm1} \rangle$ transitions producing the characteristic three peak structure visible in the excitation spectrum. The $\vert p_{\pm1} \rangle$ states are Zeeman shifted by the quadrupole field used for the MOT.}
\end{figure}

As the green probe frequency is scanned we detect the $556\,\mathrm{nm}$ fluorescence light with a photomultiplier tube (PMT). A typical excitation spectrum is shown in figure \ref{fig:typical}. The relevant levels and transitions are sketched in figure \ref{fig:levels-LS}. The three peaks visible in the signal correspond to the three $\vert g \rangle \rightarrow \vert p_{0,\pm1} \rangle$ excitation channels \cite{Maruyama2003}. The width of the $\vert g \rangle\rightarrow\vert p_0 \rangle$ excitation process (central peak) is determined by three independent effects: (i) the bare linewidth $\Gamma_p$ of the ${}^1S_0\rightarrow{}^3P_1$ transition, (ii) the Rabi broadening of the ${}^1S_0$ level induced by the presence of the $399\,\mathrm{nm}$ trapping light \cite{Loftus2000} and (iii) the Doppler broadening due to the finite sample temperature. The cloud temperature $T$ is obtained from the Gaussian width $\sigma_d$ of the Doppler contribution using the following relation:
\begin{equation}
\label{eq:T}
k_B T = \frac{1}{2} m \lambda_p^2 \sigma_d^2,
\end{equation}
where $k_B$ is the Boltzmann constant, $m$ is the atomic mass and $\lambda_p = 556\,\mathrm{nm}$ is the probe wavelength.

In order to extract the relevant information contained in the excitation spectrum we fit the central peak with a Voigt function, using the peak position and the Lorentzian and Gaussian widths as fitting parameters. The dependence of the line center on the MOT laser power, gives the light shift of the $\vert g \rangle$ level. The Lorentzian contribution is the sum of the natural linewidth and the Rabi broadening of the ground state, while the Gaussian width is given by the sample temperature. The temperature measurement is reliable only if the fitting procedure separates correctly the Gaussian and lorentzian contributions. For this reason it is important to properly describe the effect caused by the ${}^1S_0\rightarrow{}^1P_1$ MOT light onto the ${}^1S_0\rightarrow{}^3P_1$ probe and compare them with the experimental results.

We model the $\vert g \rangle\rightarrow\vert e \rangle$ transition as a four level system (taking into account the three Zeeman sublevels $\vert e_{0,\pm1} \rangle$) neglecting the presence of the probe beam. The effective Hamiltonian is therefore:
\begin{equation}
\label{eq:hamiltonian4}
H_{\mathrm{eff}}=\frac{\hbar}{2}
\left( \begin{array}{cccc}
% \begin{matrix}
 -\Delta_t		& \Omega_t/3				& \Omega_t/3		& \Omega_t/3\\
 \Omega_t^{\ast}/3	& \tilde{\Delta}_t-\Delta\omega_z'	& 0			& 0\\
 \Omega_t^{\ast}/3	& 0					& \tilde{\Delta}_t	& 0\\
 \Omega_t^{\ast}/3	& 0					& 0			& \tilde{\Delta}_t+\Delta\omega_z' \\
% \end{matrix}
\end{array}\right)
,
\end{equation}
where $\Delta_t$ and $\Omega_t$ are the trap detuning and Rabi frequency, respectively. To simplify the notation we introduce the quantity $\tilde{\Delta}_t = \Delta_t + \imath \Gamma_t$. The Zeeman splitting is given by $\Delta\omega_z'$. The Rabi frequencies on all three transition are assumed to be equal since on average an atom in the MOT sees unpolarized light. We find the ground state energy $E_g$ by diagonalizing this Hamiltonian. The imaginary part $\mathrm{Im}(E_g)$ describes the broadening of the ground state induced by laser excitation (Rabi broadening). On the other hand $\mathrm{Re}(E_g)$ accounts for the energy shift of $\vert g \rangle$ induced by the MOT beams (light shift). The agreement between the measured light shift and the result of this model is a prerequisite for the validity of the temperature measurement.

For the $\vert g \rangle\rightarrow\vert p_{\pm 1} \rangle$ transitions (side peaks) similar considerations are applied. The three effects described above contribute to the spectral line broadening in the same way. However, for these transitions the presence of the MOT magnetic field plays a crucial role. In fact the $\vert p_{\pm 1} \rangle$ states are Zeeman shifted by the trapping quadrupole field, which combined with the finite sample size gives rise to an additional line broadening. This contribution to the transition linewidth reads:
\begin{equation}
\label{eq:sigmaz}
\sigma_z = \frac{\mu_B g_p m_p b' w}{\hbar},
\end{equation}
where $\mu_B$ is the Bohr magneton, $g_p=3/2$ is the Land\'{e} factor for the ${}^1S_0\rightarrow{}^3P_1$ transition, $m_p=\pm 1$ is the magnetic quantum number, $b'$ is the MOT field gradient and $w$ is the cloud size (such that the spatial part of the atomic probability distribution is proportional to $\exp\left[-\frac{r^2}{2 w^2}\right]$). The Zeeman effect, together with gravity, is also responsible for the frequency shift of the $\vert g \rangle\rightarrow\vert p_{\pm 1} \rangle$ lines: due to imperfect MOT beam balance and alignment, the atomic cloud is formed off center with respect to the quadrupole field minimum. The resulting non-zero average magnetic field introduces an offset in the resonance frequency equal to:
\begin{equation}
\label{eq:deltaz}
\Delta\omega_z = \frac{\mu_B g_p m_p b' r_0}{\hbar},
\end{equation}
$r_0$ being the center position of the atomic cloud with respect to the quadrupole field zero.

\section{\label{sec:results1}Experimental measurement of line shift and broadenings}
As described in the previous section, the excitation spectrum obtained by probing the ${}^1S_0\rightarrow{}^3P_1$ transition carries information about the properties of the atomic cloud as well as the characteristics of the trapping potential. In the present section we apply this technique to the measurement of these quantities as a function of the MOT saturation parameter $s_t=I_t/I_{s,t}$.

During the experiment we keep the trap detuning $\Delta_t=0.86\,\Gamma_t$ and the magnetic field gradient $b'=36\,\mathrm{G/cm}$ fixed. While the atoms are trapped in the MOT the probe detuning $\Delta_p$ periodically varies from $-2\pi\times40$ to $+2\pi\times40 \, \mathrm{MHz}$ in $T=10\,\mathrm{ms}$. For each choice of $s_t$ we record the $556\,\mathrm{nm}$ fluorescence with a PMT for $1\,\mathrm{s}$ (i.e. one trace consist of $100$ spectra). Since the AOM efficiency is not uniform over the whole scanning range we divide the PMT signal by the probe power (monitored by a photodiode during the acquisition). This operation does not affect the line-shape of the spectral feature we are interested in, since the AOM efficiency loss is only significant at the borders of the scan region. However, it is crucial to have a flat background.

Once the fluorescence has been acquired and processed we extract from the full waveform the individual spectra and fit the central peak with a Voigt profile. We average the fit results and assign to each parameter an error bar equal to the calculated standard deviation.

\subsection{Ground state light shift}
\begin{figure}[h]
\includegraphics[width=0.40\textwidth]{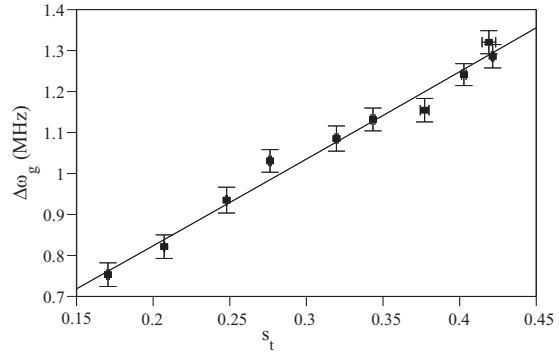}
\caption{\label{fig:lightshift1}Position of the $\vert g \rangle\rightarrow\vert p_0 \rangle$ peak relative to the resonance condition as a function of blue MOT saturation parameter $s_t$ for $\Delta_t=0.86\,\Gamma_t$. The continuous line is obtained from the four-level model presented in section \ref{sec:method}. The fitted offset and calibration factor are $\Delta\omega_{g0}=2\pi\times410\pm80\,\mathrm{kHz}$ and $\alpha=0.67\pm0.07$ respectively (see text for details).}
\end{figure}

We first focus our attention on the ground state light shift. The experimental results are shown in figure \ref{fig:lightshift1} where the $\vert g \rangle\rightarrow\vert p_0 \rangle$ peak shift $\Delta\omega_{g}$ is plotted as a function of the MOT saturation parameter. A direct comparison of the experimental data with the model presented in section \ref{sec:method} is valid only assuming that the horizontal axis calibration is not affected by systematic errors and the resonance condition is properly determined. Since we can not know {\it a priori} if these conditions are fulfilled we proceed in a different way.

We diagonalize the Hamiltonian of eq. \ref{eq:hamiltonian4} using $\Delta_t=0.86\,\Gamma_t$ and $\Delta\omega'_z=2\pi\times12.4\pm0.4\,\mathrm{MHz}$. This last value is calculated from the side peak splitting $\Delta\omega_z=2\pi\times18.6\pm0.6\,\mathrm{MHz}$ observed in the experiment by using $\Delta\omega'_z=\Delta\omega_z g_t / g_p$, being $g_t=1$ and $g_p=3/2$ the Land\'{e} factors for the trap and probe transition respectively. We fit the experimental data for $\Delta\omega_g$ with $\mathrm{Re}(E_g)/\hbar$ derived from the diagonalization. As fitting parameters we use the calibration factor $\alpha$ (i.e. the saturation parameter used in $H$ is $\alpha s_t$) and the offset $\Delta\omega_{g0}$. We obtain $\alpha=0.67\pm0.07$ and $\Delta\omega_{g0}=2\pi\times410\pm80\,\mathrm{kHz}$. The calculated behaviour is represented in figure \ref{fig:lightshift1} as a continuous line.

The fact that $\Delta\omega_{g0}$ is different from zero indicates that the $556\,\mathrm{nm}$ laser source may be locked with non-zero detuning with respect to the atomic resonance. However it also has to be considered that during the measurement process the Zeeman slower beam is constantly present thus contributing to the ground state light shift. With only this measurement we cannot distinguish between the two effects.

Concerning the value of $\alpha$ it is important to notice that the saturation parameter calibration assumes that the atomic cloud is formed at the intersection of the six trapping beams, considered to be perfectly Gaussian. It is realistic to assume that the actual intensity at the MOT position is lower than the estimated value. Therefore the measured mismatch of about $30\%$ can be considered to be in reasonable agreement with our expectation.

\subsection{\label{subsec:ex}Mapping the trap laser intensity: an example}
The observed value of $\alpha$ gives us information about the actual light intensity at the MOT position. This suggests that the trapped atoms can be used as a probe to explore the spatial distribution of the light field for a given MOT alignment.

The measurement is performed as follows. First we load the ${}^1S_0\rightarrow{}^1P_1$ MOT and we probe the ${}^1S_0\rightarrow{}^3P_1$ transition recording the excitation spectrum. We then shift the quadrupole field zero by applying a uniform magnetic field and we perform a new fluorescence measurement. The sequence is repeated for different trap positions while the laser beam alignment remains unchanged. Two pairs of Helmholtz coils generate a uniform magnetic field along the vertical ($\hat{x}$) and Zeeman slower ($\hat{y}$) directions of $8.6\,\mathrm{G/A}$ and $4.1\,\mathrm{G/A}$ respectively. For the typical gradient of $36\,\mathrm{G/cm}$ ($18\,\mathrm{G/cm}$ along $\hat{y}$) used in the experiment the corresponding position shifts are $0.28\,\mathrm{mm/G}$ along $\hat{x}$ and $0.56\,\mathrm{mm/G}$ in the $\hat{y}$ direction.

\begin{figure}[h]
\includegraphics[width=0.45\textwidth]{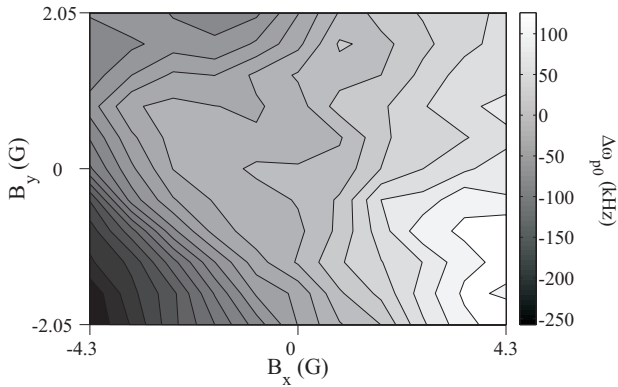}
\caption{\label{fig:mapping} Position-dependent ground state light shift ameasured through variation of compensation coil currents. The frequency shift is referred to its value in absence of compensation field. Laser intensity and detuning are maintained fixed at $s_t=0.5$ and $\Delta_t = 0.86\,\Gamma_t$.}
\end{figure}

During the mapping sequence we vary both coil currents from $-0.5$ to $0.5\,\mathrm{A}$ in steps of $0.1\,\mathrm{A}$. For each current configuration we acquire $10$ excitation spectra and we analyze them as described above. In figure \ref{fig:mapping} we plot the frequency offset of the $\vert g \rangle\rightarrow\vert p_0 \rangle$ peak as a function of the offset fields. Here $\Delta\omega_{g}$ is referenced to its value in absence of external fields (i.e. $\Delta\omega_{g}=0$ for $B_x,B_y=0$). During the experimental sequence the MOT light intensity was kept constant at $s_t=0.5$. Therefore the observed light shift variation is only due to the displacement of the MOT across the intensity distribution of the trapping light field.

We observe that the light shift increases with $B_x$ (i.e. as the atoms are moved downwards). Therefore the light field maximum is located below the cloud position for $B_x=0$. Since we aligned the MOT for $B_x=B_y=0$ maximizing the number of atoms, we conclude that for the optimum alignment the cloud is formed above the intersection of the trapping beam. We also observed that the optimum MOT loading is achieved when the Zeeman slower beam passes below the atom position. An interpretation of this result would require a theoretical treatment which is beyond the purpose of this paper.

The experiment presented here is an example of how the technique discussed in this paper can be used to measure the properties of the trapping potential. For this particular measurement we observe the $\vert g \rangle\rightarrow\vert p_0 \rangle$ peak shift and we relate it to the local trap field intensity. In the following we will show how a proper determination of the Lorentzian and Gaussian contributions to the peak linewidth, can provide a reliable measurement of the sample temperature. We think that this technique could be helpful for unveiling the actual cloud dynamics.

\subsection{Rabi broadening}

\begin{figure}[h]
\includegraphics[width=0.45\textwidth]{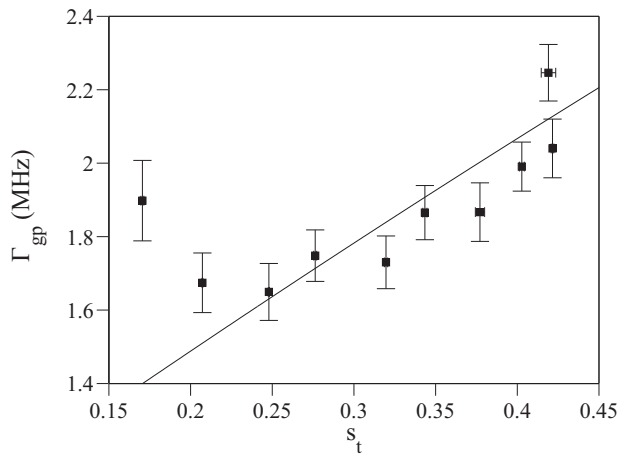}
\caption{\label{fig:lightshift2}Central peak Lorentzian width $\Gamma_{gp}$ as a function of MOT saturation parameter $s_t$. The continuous line is obtained diagonalizing the Hamiltonian of eq. \ref{eq:hamiltonian4} with $\alpha=0.67$. The minimization process gives an offset $\Gamma_{gp0}=2\pi\times800\pm60\,\mathrm{kHz}$ (see text for details).}
\end{figure}

The presence of $399\,\mathrm{nm}$ MOT light also modifies the linewidth of the ground state $\vert g \rangle$ (Rabi broadening). As a result the Lorentzian contribution to the central peak width ($\Gamma_{gp}$) increases as a function of the trap saturation parameter $s_t$. In figure \ref{fig:lightshift2} we plot the experimental values obtained from the data set of figure \ref{fig:lightshift1}. The comparison with the four-level model is carried out following the same procedure as for the ground state light shift analysis. In this case the Hamiltonian of eq. \ref{eq:hamiltonian4} is diagonalized assuming that $\alpha=0.67$. The only free parameter is therefore $\Gamma_{gp0}$, i.e. the value of $\Gamma_{gp}$ in the limit of zero intensity. A minimization process similar to the one applied before leads to $\Gamma_{gp0}=2\pi\times800\pm60\,\mathrm{kHz}$. The fit is plotted in figure \ref{fig:lightshift2} as a continuous line.

As discussed in section \ref{sec:method} the Lorentzian width of the $\vert g \rangle \rightarrow \vert p_0 \rangle$ peak is also determined by the bare atomic transition linewidth. Therefore for $s_t=0$ we expect to have $\Gamma_{gp0} = \tilde{\Gamma}_p$, where $\tilde{\Gamma}_p=\Gamma_p\sqrt{1+s_p}$. From the probe beams size and power we have $s_p \approx 9$, therefore $\tilde{\Gamma}_p \approx 2\pi\times580\,\mathrm{kHz}$. The difference between $\Gamma_{gp0}$ and $\tilde{\Gamma}_p$ is attributed to the presence of the Zeeman slower beam during the measurement and to the linewidth of our probe laser ($2\pi\times10\,\mathrm{kHz}$). For $s_t>0.25$ the experimental data show good agreement with the fit. The deviation observed for $s_t<0.25$ is due to the fact that the Voigt fit is not accurate when the Lorentzian contribution to the line profile is too small.

\subsection{Cloud temperature}
\begin{figure}[h]
\includegraphics[width=0.45\textwidth]{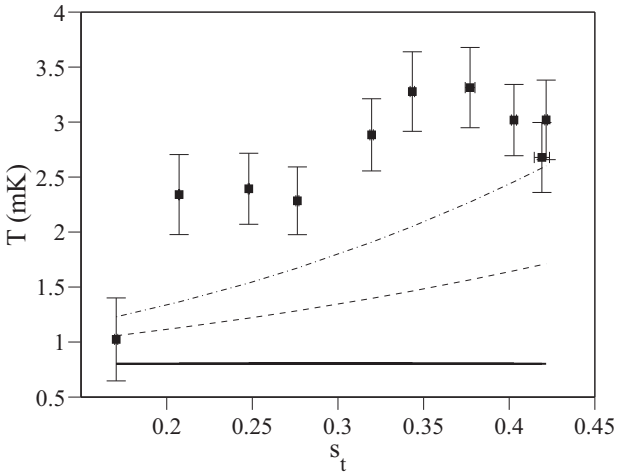}
\caption{\label{fig:temp1}Cloud temperature as a function of $s_t$. The continuous curve is the standard Doppler limit. the dot-dashed and dashed line are obtained from the theoretical model discussed in \cite{Choi2008} calculated with $\alpha=1$ and $\alpha=0.67$ respectively (see text for details).}
\end{figure}

The analysis described above shows that both light shift and Rabi broadening are consistently taken into account and extracted from the fluorescence profiles. The residual Gaussian width can now be related to the cloud temperature via eq. \ref{eq:T}. Experimental data are plotted in figure \ref{fig:temp1}. The continous line represent the standard Doppler limit $T_d=695\,\mathrm{\mu K}$ which is independent of the trap light intensity. The dot-dashed line is derived from the model presented in the paper by Choi et al. (eq. 13 of \cite{Choi2008}). The dashed line is obtained in the same way but replacing the experimental value of $s_t$ with $\alpha s_t$ being $\alpha=0.67$ the calibration factor derived from our previous analysis.

We observe that the measured temperature is higher than the standard Doppler limit as predicted by the model of reference \cite{Choi2008}. However experimental data are systematically above the theoretical curve and when the calibration factor is applied the deviation between experiment and theory increases. The lack of quantitative agreement with the experiment could be related to the presence of other effects not included in the model. In particular, as shown in \cite{Chaneliere2005}, spatial intensity fluctuations of the cooling beams are responsible for extra-heating. From this perspective the knowledge about the actual trap light intensity at the cloud position observed in section \ref{subsec:ex}, could be a relevant ingredient for a proper theoretical treatment of the cooling dynamics of two-electrons atoms.

\section{\label{sec:results2}Time resolved measurement}
As discussed in section \ref{sec:results1} the presence of the Zeeman slower beam during the measurement process and the detuning of the probe laser with respect to the ${}^1S_0\rightarrow{}^3P_1$ transition influence the determination of the ground state light shift. In order to avoid these effects we proceed as follows.

As a first step we perform a fluorescence measurement on trapped atoms as described in the previous section. We then switch off the blue light by means of an AOM. The atoms decay in a time $\tau_t=1/\Gamma_t=5.5\,\mathrm{ns}$ to the  $\vert{}^1S_0\rangle$ state. Since the ground level has $0$ angular momentum, the MOT gradient does not influence the atomic motion. Finally we perform a second fluorescence measurement on free falling atoms. 

The absence of confinement during the last stage forces us to realize the measurement faster than the time needed for the atoms to leave the probe region. For this reason we keep the frequency span at $2\pi\times80\,\mathrm{MHz}$ as in section \ref{sec:results1} and reduce the scan period to $1\,\mathrm{ms}$. Note that during this time the cloud's center-of-mass moves by about $5\,\mathrm{\mu m}$. With the resulting scan rate of $R=2\pi\times80\,\mathrm{MHz/ms}$ the probe frequency stays on resonance for a time equal to $\tau_s=\Gamma_p / R = 2.3\,\mathrm{\mu s}$. This value has to be compared to the atomic decay time $\tau_p=870\,\mathrm{ns}$. The fact that $\tau_s \approx 2.6 \tau_p$ ensures that the fluorescence signal is not affected by non-adiabaticity.

\begin{figure}[h]
\includegraphics[width=0.45\textwidth]{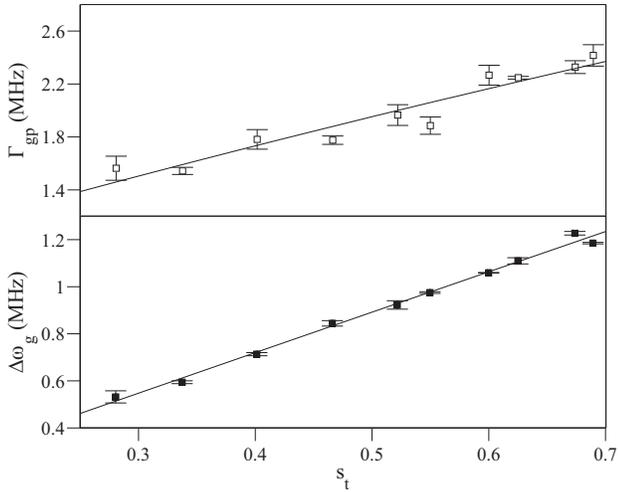}
\caption{\label{fig:lightshift3}Rabi broadening (top) and light shift (bottom) as a function of $s_t$ (vertical axis unit is $\mathrm{MHz}$). The experimental data are obtained comparing spectroscopy in the presence and absence of $399\,\mathrm{nm}$ trapping light. The continuous lines are obtained from the four-level model of section \ref{sec:method}. The trap detuning is $\Delta_t=0.86\,\Gamma_t$. The magnetic field gradient is $36\,\mathrm{G/cm}$.}
\end{figure}

The difference between the $\vert g \rangle \rightarrow \vert p_0 \rangle$ peak positions with and without trapping light is only related to the ground state light shift caused by the MOT beams. The experimental results are shown in figure \ref{fig:lightshift3} (bottom) together with the model comparison (see section \ref{sec:results1}). The minimization process gives us $\alpha=0.55\pm0.04$ and $\Delta\omega_{g0}=2\pi\times30\pm60\,\mathrm{kHz}$. The calibration factor $\alpha$ is in agreement with the previously measured value. The offset $\Delta\omega_{g0}$ is compatible with zero as expected.

From the same data set we obtain the ground state Rabi broadening as described in section \ref{sec:results1}. For this purpose we use the spectra acquired in the presence of $399\,\mathrm{nm}$ light. Experiment and model are shown in figure \ref{fig:lightshift3} (top). As in section \ref{sec:results1}, $\alpha$ is kept fixed at the previously obtained value of $0.55$ while for $\Gamma_{gp0}$ we find $2\pi\times800\pm30\,\mathrm{kHz}$.

\begin{figure}[h]
\includegraphics[width=0.45\textwidth]{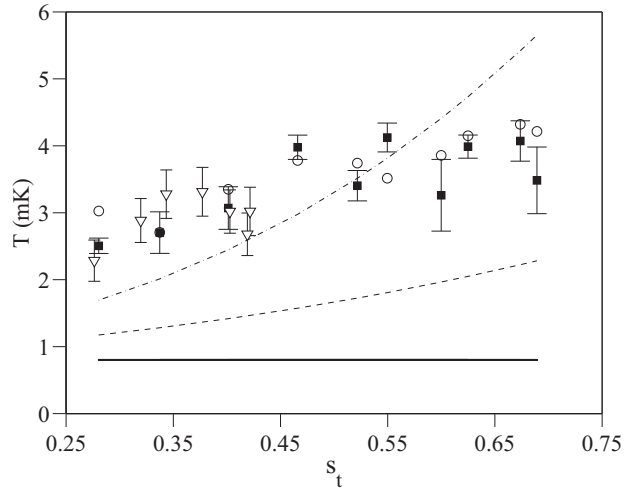}
\caption{\label{fig:temp2}Cloud temperature measured with (empty circles) and without (full squares) MOT light. For the empty circles the experimental errors are smaller that the symbol size. The data of figure \ref{fig:temp2} are also shown (empty triangles). The continous line is the standard Doppler limit. The prediction given by the model of \cite{Choi2008} is plotted as a dash-dotted line (dashed with calibration correction).}
\end{figure}

In figure \ref{fig:temp2} we plot the cloud temperature calculated from eq. \ref{eq:T} using the fluorescence signal measured before (empty circles) and after (full squares) switching off the blue MOT. In the latter case the spectra are fitted with a fixed Lorentzian width equal to the value of $\Gamma_{gp0}$ obtained from our previous analysis. The two sets are compatible showing that the Voigt fit correctly identifies the Gaussian component independently of the presence of the trapping light. As for figure \ref{fig:temp1} we plot the standard Doppler limit (continuous) and the theoretical curve after reference  \cite{Choi2008} calculated for $\alpha=1$ (dot-dashed) and $\alpha=0.55$ (dashed line).

As detailed above, each excitation spectrum is acquired in $T=1\,\mathrm{ms}$. This value is $10$ times smaller than the one used for the experiments described in section \ref{sec:results1}. However, if the scan is adiabatic, the temperature measurement does not depend on the particular choice of $T$. In order to verify this assumption, we reproduce in figure \ref{fig:temp2} the experimental points previously presented in figure \ref{fig:temp1} (empty triangles). We note that, whitin the experimental error, the three data sets.

Due to the relatively short scan period $T$ used in the experiment, we regard the information contained in each excitation spectrum as ``instantaneous''. In other words we are performing a time resolved measurement of the atomic cloud properties. This suggest a different way of comparing the temperature obtained using the technique of \cite{Honda1999,Loftus2000a} with the result of a more standard time-of-flight (TOF) approach. The procedure we use is described as follows.

We first load the $399\,\mathrm{nm}$ MOT and then we switch off the trapping beams. After a variable time $t$ we perform a single probe frequency scan in $1\,\mathrm{ms}$. We fit the three peaks present in the fluorescence signal with Voigt profiles. We then calculate the difference between the Gaussian widths of the central and side peaks. In this way we obtain the contribution $\sigma_z$ to the side peak broadening due to the presence of the MOT magnetic field gradient. Finally $\sigma_z$ can be related to the cloud size via equation \ref{eq:sigmaz}. This procedure resembles the standard TOF technique. The only difference between the two approaches resides in the way used to determine the sample size: in one case by absorption/fluorescence imaging (direct), in the other one by excitation spectroscopy (indirect).

\begin{figure}[h]
\includegraphics[width=0.45\textwidth]{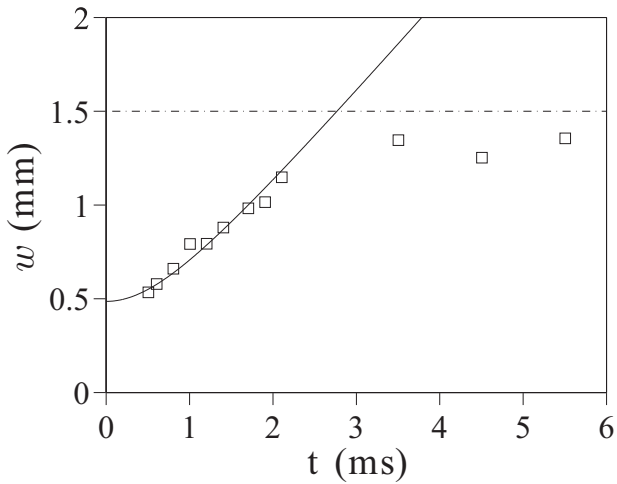}
\caption{\label{fig:TOF}Cloud size $w$ as a function of TOF $t$. The horizontal dot-dashed line represent the detector aperture. The continuous line is the obtained fitting the experimental data with the function of eq. \ref{eq:wt}.}
\end{figure}

In figure \ref{fig:TOF} we plot the cloud size $w$ as a function of the free fall time $t$. As it is clearly visible in the graph $w$ is constant for $t>2.5\,\mathrm{ms}$. This effect can be easily understood considering that as $t$ increases the cloud becames bigger and eventually its image on the PMT exceedes the detector aperture. When this occurs only part of the atomic distribution contributes to the observed fluorescence signal. In figure \ref{fig:temp2} we place a dot-dashed horizontal line at a position equal to one half of the PMT aperture radius. When the actual cloud size is equal to this value, $5\%$ of the atomic density ditribution (defined as in section \ref{sec:method}) is not imaged onto the detector.

We fit the data points for $t<2.5\,\mathrm{ms}$ with the expected expression for free ballistic expansion:
\begin{equation}
\label{eq:wt}
w(t) = w(0) \sqrt{1+\frac{k_b T}{m} t^2},
\end{equation}
where $m$ is the atomic mass and $w(0)$ is the initial cloud size. From the fit we get $T=5.5\pm1.2\,\mathrm{mK}$. This result has to be compared with the correspondent value determined by a fully spectroscopic method. For this purposes we analyze the spectra acquired before releasing the trap following the same procedure as in section \ref{sec:results1}. This procedure leads to $T=4.91\pm0.06\,\mathrm{mK}$ which is in good agreement with the TOF value.

\section{\label{sec:conclusions}Conclusions}
We extended the method proposed in \cite{Honda1999,Loftus2000a} for online monitoring of the temperature of cold atoms. Probing the ${}^1S_0\rightarrow{}^3P_1$ transition of ${}^{174}\mathrm{Yb}$ atoms confined in a magneto-optical trap we measured the ground state light shift and broadening induced by the MOT light field. We then fit the experimental data with a simple model describing the ${}^1S_0\rightarrow{}^1P_1$ transition. As a results we calibrated the MOT local intensity. This allowed us to extract the Gaussian component of the excitation spectra and measure the sample temperature avoiding the limitations encountered by the authors of \cite{Loftus2000a}.

We applied a similar procedure to free-falling atoms. We compared the measurements performed with and without confining potential thus obtaining the ground state light shift in absence of spurious effects (like additional light shifts and laser locking offset). We showed that the temperature measurement is not affected by $399\,\mathrm{nm}$ light indicating that the Gaussian contribution to the line is properly determined. Finally we measure the sample temperature from free cloud expansion. In contrast with traditional time-of-flight measurement we used the $\vert g \rangle\rightarrow\vert p_{\pm 1} \rangle$ linewidth for determining the cloud size. We showed that the results obtained with the two method are in agreement.

The measured temperature of our $\mathrm{Yb}$ MOT is consistently above the Doppler limit. This result agrees with the observations performed by other groups \cite{Xu2002,Xu2003}. We confirmed that the theoretical model presented in \cite{Choi2008} qualitatively describes the dependence of the cloud temperature on the trap light intensity. However our data do not show a quantitative agreement with the theory. We measured the actual MOT light intensity distribution at the cloud position. We pointed out that this could account for the observed disagreement.

Providing that the adiabaticity condition is satified, the acquisition time $T$ can be reduced without affecting the information carried by the excitation spectrum. The value of $T=1\,\mathrm{ms}$ used in the experiment, does not have to be regarded as a fundamental limit. Keeping the scan rate $R$ fixed, $T$ can be lowered by choosing a different frequency span. Thanks to this fact the cloud temperature and size could be monitor on a fast time scale.

In conclusion the technique presented in this paper opens the possibility to monitor online and in ``real time'' the dynamics of a cold atomic cloud. This could be a key tool to understand the peculiar thermodynamics of magneto-optical trapping of two-eletrons atoms.

\section{\label{sec:acknowledgments}Acknowledgments}
The Authors thank professor E. Arimondo for helpful discussions. M. Cristiani and T. Valenzuela gratefully acknowledge the Juan de la Cierva fellowship by MICINN within plan nacional I+D+I 2004-2007. This work is supported by Consolider-Ingenio project ``QOIT'' (CSD2006-00019) and by the EFS through the individual project CMMC within the EuroQUAM collaborative research project.

\bibliography{TemPap-springer}

\end{document}